\documentclass[lettersize,journal]{IEEEtran}
\usepackage[skip=4pt]{caption}
\usepackage{amsmath,amsfonts}
\usepackage{algorithmic}
\usepackage{threeparttable}
\usepackage{algorithm}
\usepackage{array}
\usepackage[caption=false,font=normalsize,labelfont=sf,textfont=sf]{subfig}
\usepackage{textcomp}
\usepackage{stfloats}
\usepackage{url}
\usepackage{multirow}
\usepackage{verbatim}
\usepackage{graphicx}
\usepackage{cite}
\usepackage{xcolor}
\usepackage{tikz}
\usepackage{booktabs}
\usepackage[utf8]{inputenc}
\usepackage{amssymb}
\usepackage{bm}
\usetikzlibrary{arrows.meta,positioning,fit,calc,backgrounds}
\definecolor{upd}{RGB}{213,94,0}
\definecolor{frz}{gray}{0.45}
\definecolor{upd}{RGB}{0,114,178}      
\definecolor{updfill}{RGB}{218,236,248} 
\usetikzlibrary{positioning,calc,arrows.meta}
\begin{document}

\definecolor{frozen}{HTML}{6B7280}     
\definecolor{frozenbg}{HTML}{F1F2F4}  

\definecolor{learn}{HTML}{2E4374}       
\definecolor{learnbg}{HTML}{DCE3F5}     

\definecolor{actln}{HTML}{A3A3A3}      
\definecolor{actbg}{HTML}{FBFBFB}    
\title{Meta-Transfer Learning for mmWave Beam Alignment}

\author{
A.~Nuri Cevik and Sinem Coleri

\thanks{This work has been submitted to the IEEE for possible publication. Copyright may be transferred without notice, after which this version may no longer be accessible.}
\thanks{A.N.~Cevik was with the Department of Electrical and Electronics
Engineering, Koc University, 34450 Istanbul, T\"urkiye, and is now with
the Laboratory for Information and Decision Systems (LIDS),
Massachusetts Institute of Technology, Cambridge, MA 02139 USA
(e-mail: ancevik@mit.edu).}
\thanks{S.~Coleri is with the Department of Electrical and Electronics
Engineering, Koc University, 34450 Istanbul, T\"urkiye
(e-mail: scoleri@ku.edu.tr).}
}

\markboth{}{Cevik and Coleri: Meta-Transfer Learning for mmWave Beam Alignment}

\maketitle

\begin{abstract}
Millimeter-wave (mmWave) beam alignment is critical for next-generation
wireless systems, but existing approaches either update all parameters
during adaptation or restrict updates to a subset of layers without adapting
intermediate feature representations. We propose MTL-BA, a meta-transfer
learning framework for multiple-input single-output (MISO) beam alignment
that freezes a pre-trained convolutional backbone and meta-learns lightweight
Scale-and-Shift (SS) adapters and a classifier head. Under an outdoor-to-indoor shift in DeepMIMO, MTL-BA achieves the highest Top-1 accuracy among the evaluated methods at 35~dB while updating $17\times$ fewer parameters than full fine-tuning and meta-training $6\times$ faster than MAML.
\end{abstract}

\section{Introduction}
\IEEEPARstart{M}{illimeter-wave} (mmWave) communication is a cornerstone of fifth-generation (5G) and beyond wireless systems due to its multi-gigahertz bandwidth. However, its high path loss and blockage sensitivity necessitate large antenna arrays at the base station (BS), demanding precise beam alignment with each user equipment (UE). Conventional exhaustive search over a narrow-beam codebook incurs sweeping overhead that scales linearly with codebook size, becoming prohibitive for large arrays.

To mitigate this overhead, deep learning (DL)-based beam 
prediction methods have been extensively investigated. A 
prominent approach replaces exhaustive sweeping with a small 
set of wide probing beams drawn from a DFT codebook: the 
received power measurements are fed into a neural network 
that directly predicts the optimal narrow beam from an 
oversampled DFT (O-DFT) codebook, significantly reducing 
beam training overhead~\cite{heng2022, khan2025}. However, such methods assume identical training and deployment distributions, whereas channel characteristics vary across sites, carrier frequencies, and propagation environments, degrading a statically trained model.

A practical remedy is to adapt a model pre-trained at existing deployment sites
to a new site using a small set of labeled samples, e.g., collected via
occasional exhaustive sweeps. Making this adaptation efficient in both samples
and computation is a central challenge. Existing approaches fall into two
categories. Full-network meta-learning methods~\cite{yuan2021tl_ml, samba2023}
train a convolutional neural network (CNN) from random initialization via
episodic optimization across multiple channel distributions, enabling rapid
adaptation but updating all network weights and requiring costly
meta-training. Partial-adaptation methods~\cite{yuan2021tl_ml,
zhang2022embedding, yang2023dualband, mu2025mpba} reduce adaptation cost by
restricting updates to a subset of layers, such as the classifier head or
decoder, but do not modulate the intermediate feature representations to the
target domain and, when combined with meta-learning, still require training
from random initialization.

In this letter, we propose MTL-BA, a meta-transfer learning framework that
integrates a frozen pre-trained CNN backbone with lightweight Scale-and-Shift
(SS) adapters.
(i)~MTL-BA meta-learns only the SS adapters and classifier head via first-order
episodic training. Since the beam codebook, and hence the label space, is shared
across sites, the pre-trained head is retained rather than discarded as in
few-shot meta-transfer learning~\cite{sun2019mtl}.
(ii)~Under an outdoor 28~GHz to indoor 60~GHz shift in DeepMIMO~\cite{deepmimo},
we compare MTL-BA with full fine-tuning, low-rank adaptation
(LoRA)~\cite{hu2022lora}, normalization-based adaptation~\cite{li2018adabn},
and model-agnostic meta-learning (MAML)~\cite{finn2017maml} in accuracy, spectral efficiency, and training cost.
(iii)~We show that meta-training and SS adaptation are complementary, dominating
the gain when adaptation samples are scarce and abundant, respectively.

The remainder of this letter is organized as follows. 
Section~II presents the system model. Section~III formulates 
the beam prediction problem and details the proposed MTL-BA 
framework. Section~IV provides simulation results. Section~V 
concludes the letter.
\section{System Model}
We consider a mmWave multiple-input single-output (MISO) system where the BS is
equipped with a uniform linear array (ULA) with $N_{\mathrm{BS}}$ antennas and
serves $N_U$ single-antenna UEs, each using a single stream. The channel from
the BS to UE$_u$ can be expressed based on geometric channel modeling as
\begin{equation}
\mathbf{h}_u=\sum_{l=1}^{L}\alpha_{u,l}\,\mathbf{b}(\phi_{u,l}),
\end{equation}
where $L$ denotes the number of paths, $\alpha_{u,l}$ is the complex path gain
of the $l$-th path, $\phi_{u,l}$ is its angle of departure, and
$\mathbf{b}(\phi)=\frac{1}{\sqrt{N_{\mathrm{BS}}}}\big[1,e^{j\psi},\dots,
e^{j(N_{\mathrm{BS}}-1)\psi}\big]^{T}$ is the array response vector with
$\psi=\pi\sin\phi$ for half-wavelength antenna spacing.

We consider an analog beamformer in the radio frequency (RF) domain, in which
each antenna element is connected to a phase shifter, so that every beamformer
satisfies the power and constant modulus constraints. During beam sweeping, the
BS transmits a unit-power symbol $s$ using a beam $\mathbf{v}_i$ selected from
a predefined codebook $\mathbf{V}\in\mathbb{C}^{N_{\mathrm{BS}}\times N_V}$
with $N_V$ candidate beams. The received signal at UE$_u$ is
\begin{equation}
r_{u,i}=\sqrt{P_T}\,\mathbf{h}_u^{H}\mathbf{v}_i\,s+n_u,
\end{equation}
where \(P_T\) denotes the transmit power and
\(n_u\sim\mathcal{CN}(0,\sigma_n^2)\) denotes circularly symmetric
complex Gaussian noise with \(\mathbb{E}[|n_u|^2]=\sigma_n^2\).
The resulting SNR is
\(\mathrm{SNR}_{u,i}=P_T|\mathbf{h}_u^{H}\mathbf{v}_i|^2/\sigma_n^2\),
and the optimal beam index is
\begin{equation}
i_u^{*}=\arg\max_{i\in\{1,\dots,N_V\}}\mathrm{SNR}_{u,i}
=\arg\max_{i\in\{1,\dots,N_V\}}\big|\mathbf{h}_u^{H}\mathbf{v}_i\big|^2.
\label{eq:opt_beam}
\end{equation}
\section{Deep Neural Network Framework for Beam Alignment}
We first formulate DL-based beam prediction and then present MTL-BA, which freezes a pre-trained CNN backbone and meta-learns lightweight SS adapters for adaptation under distribution shift.
\subsection{DL-based Beam Prediction Formulation}
Instead of sweeping the entire narrow-beam codebook, the BS sequentially
transmits pilot symbols over \(M_p \ll N_V\) wide probing beams, each in a
separate time slot. For each user \(u\), the received signal corresponding to
the \(p\)-th probing beam is given by
\begin{equation}
r_{u,p} = \sqrt{P_T}\,\mathbf{h}_u^H \mathbf{w}_p\, s_p + n_{u,p},
\qquad p = 1,\dots,M_p,
\end{equation}

where \(\mathbf{w}_p\) denotes the \(p\)-th probing beam, \(s_p\) is the
transmitted pilot symbol, and \(n_{u,p}\) is noise with the same statistics as
in~(2). Each UE measures the received signal strength over the \(M_p\) probing
beams and forms the input feature vector
\begin{equation}
\mathbf{x}_u = \left[|r_{u,1}|^2,\; |r_{u,2}|^2,\; \cdots,\; |r_{u,M_p}|^2\right]^T.
\label{eq:probe_vec}
\end{equation}
It is assumed that the probing, measurement, and feedback operations are
completed within the channel coherence time.

We formulate beam prediction as a supervised classification task. A parametric
function \(f(\cdot;\boldsymbol{\omega}): \mathbb{R}^{M_p} \rightarrow
\mathbb{R}^{N_V}\) with learnable parameters \(\boldsymbol{\omega}\) maps
\(\mathbf{x}_u\) to a probability vector \(\hat{\mathbf{q}}_u =
f(\mathbf{x}_u;\boldsymbol{\omega})\), whose \(i\)-th element
\(\hat{q}_{u,i}\) is the predicted probability that beam \(i\) is optimal. The
parameters are trained by minimizing the cross-entropy with respect to the
optimal beam index in~(3),
\begin{equation}
\min_{\boldsymbol{\omega}} \; \mathbb{E}_{u}\!\left[-\log \hat{q}_{u,i_u^*}\right],
\end{equation}
and the predicted beam is \(\hat{i}_u = \arg\max_{i} \hat{q}_{u,i}\). The formulation extends to single-stream MIMO by treating transmit--receive
beam pairs as prediction classes and using the corresponding probing
measurements as input.


\subsection{Meta-Transfer Learning for Beam Alignment}
\label{subsec:mtl}

To enable fast adaptation to unseen environments using only a small number of
labeled samples, we adopt a meta-transfer learning (MTL) strategy inspired by
Sun et al.~\cite{sun2019mtl}, originally proposed for few-shot image
classification. The key idea is to freeze a pre-trained backbone and meta-learn
a set of lightweight Scale-and-Shift (SS) parameters
$\boldsymbol{\Phi}=\{\boldsymbol{\phi}^{\gamma},\boldsymbol{\phi}^{\beta}\}$ that
modulate the frozen feature representations, together with a task-adaptive
classifier head. We partition the pre-trained parameters as
$\boldsymbol{\omega}=\{\boldsymbol{\Theta},\boldsymbol{\theta}\}$, where
$\boldsymbol{\Theta}$ denotes the backbone and $\boldsymbol{\theta}$ the
classifier head. The SS parameters are newly introduced, and both
$\boldsymbol{\Phi}$ and $\boldsymbol{\theta}$ are meta-learned through episodic
training across multiple source BS environments, as illustrated in
Fig.~\ref{fig:mtlba}.

\subsubsection{Model Architecture with Scale-Shift Adapters}
The backbone weights \(\boldsymbol{\Theta}\) are kept frozen. SS operations are inserted after the activation of each convolutional layer and of the first fully connected layer. Each applies a channel-wise (neuron-wise for the fully connected layer) affine transformation to the feature tensor \(\mathbf{z}\) as
\begin{equation}
\mathrm{SS}(\mathbf{z}\,;\boldsymbol{\phi}^{\gamma},\boldsymbol{\phi}^{\beta})
  = \boldsymbol{\phi}^{\gamma} \odot \mathbf{z} + \boldsymbol{\phi}^{\beta},
\label{eq:ss_op}
\end{equation}
where \(\odot\) denotes element-wise multiplication. Initializing \(\boldsymbol{\phi}^{\gamma}=\mathbf{1}\) and \(\boldsymbol{\phi}^{\beta}=\mathbf{0}\) makes the adapted network initially coincide with the pre-trained one, so the pre-trained weights are preserved.

\begin{figure}[!t]
\centering
\begin{tikzpicture}[
  font=\footnotesize,
  >={Stealth[length=3.5pt]},
  box/.style={draw, rounded corners=1.5pt, minimum height=6.5mm, align=center, inner xsep=2pt, inner ysep=2.5pt},
  data/.style={box, draw=black!55, fill=white},
  frozen/.style={box, draw=frz, fill=black!8},
  updated/.style={box, draw=upd, fill=updfill},
  arr/.style={->, draw=black!70, semithick},
  panel/.style={draw=black!35, rounded corners=3pt, inner xsep=2mm, inner ysep=2.2mm},
  title/.style={font=\footnotesize\bfseries, fill=white, inner sep=1pt},
  note/.style={font=\scriptsize, text=black!70, align=center},
]

\node[updated, minimum width=26mm] (inner) at (-23mm,0) {Inner loop \eqref{eq:inner_update}\\[1pt]$\bm{\theta}\to\bm{\theta}_k^{(G_{\mathrm{in}})}$};
\node[updated, minimum width=26mm] (outer) at ( 23mm,0) {Outer loop \eqref{eq:outer_ss}, \eqref{eq:outer_head}\\[1pt]$\bm{\Phi},\bm{\theta}$};
\draw[arr] (inner) -- (outer);
\node[data, above=4mm of inner, minimum height=5mm] (sup) {$\mathcal{S}_k$};
\node[data, above=4mm of outer, minimum height=5mm] (qry) {$\mathcal{Q}_k$};
\draw[arr] (sup) -- (inner); \draw[arr] (qry) -- (outer);
\node[note] (task) at (0,0 |- sup) {$\mathcal{T}_k$ from $b_k\in\mathcal{B}$};
\draw[arr] (outer.south) -- ++(0,-3mm) -| (inner.south);
\node[note, below=3mm of $(inner.south)!0.5!(outer.south)$, anchor=north, inner sep=0.5pt] (nx) {next episode};

\matrix[column sep=2.5mm, anchor=south] (A) at ($(0,0 |- sup.north)+(0,10mm)$) {
  \node[data] (src) {Source sites $\mathcal{B}$}; & \node[updated] (cnn) {CNN $\bm{\Theta},\bm{\theta}$}; \\
};
\draw[arr] (src) -- (cnn);

\matrix[column sep=2.5mm, anchor=north] (C) at ($(0,0 |- nx.south)+(0,-10mm)$) {
  \node[data] (dad) {$\mathcal{D}_{\mathrm{ad}}$}; & \node[frozen] (th) {$\bm{\Theta}$}; &[-2.5mm]
  \node[updated] (pt) {$\bm{\Phi},\bm{\theta}$}; & \node[data] (out) {$\hat{i}_u$}; \\
};
\draw[arr] (dad) -- (th); \draw[arr] (pt) -- (out);

\coordinate (L) at ($(inner.west)+(-1mm,0)$);
\coordinate (R) at ($(outer.east)+(1mm,0)$);
\begin{scope}[on background layer]
  \node[panel, fit=(A)(L|-A)(R|-A)] (P1) {};
  \node[panel, fit=(inner)(outer)(sup)(qry)(nx)(L|-inner)(R|-inner)] (P2) {};
  \node[panel, fit=(C)(L|-C)(R|-C)] (P3) {};
\end{scope}

\node[title, anchor=west] at ($(P1.north west)+(2.5mm,0)$) {(a) Pre-training};
\node[title, anchor=west] at ($(P2.north west)+(2.5mm,0)$) {(b) Meta-training};
\node[title, anchor=west] at ($(P3.north west)+(2.5mm,0)$) {(c) Target adaptation \eqref{eq:adaptation}};
 
\draw[arr, thick] (P1.south) -- node[right, note] {$\bm{\Theta},\bm{\theta}$} (P2.north);
\draw[arr, thick] (P2.south) -- node[right, note] {$\bm{\Phi},\bm{\theta}$} (P3.north);
\end{tikzpicture}
\caption{Overview of MTL-BA. Blue blocks are updated and gray blocks are frozen.}
\label{fig:mtlba}
\end{figure}

\subsubsection{Episodic Training}
Let \(\mathcal{B}=\{b_1,\dots,b_K\}\) be the set of \(K\) source BS environments used for meta-training. Each epoch consists of \(T_{\mathrm{ep}}\) episodes. In each episode, a task \(\mathcal{T}_k\) is sampled from environment \(b_k\in\mathcal{B}\), and its data are partitioned into a support set \(\mathcal{S}_k=\{(\mathbf{x}_u,i_u^{*})\}_{u=1}^{N_s}\) and a query set \(\mathcal{Q}_k=\{(\mathbf{x}_u,i_u^{*})\}_{u=1}^{N_q}\). Similar to~\cite{sun2019mtl}, tasks are processed one at a time, and the outer update uses the query loss of a single task rather than the average over a meta-batch as in MAML~\cite{finn2017maml}. The loss over a data set \(\mathcal{D}\) is the cross-entropy
\begin{equation}
\mathcal{L}_{\mathcal{D}}(\boldsymbol{\theta},\boldsymbol{\Phi};\boldsymbol{\Theta})
= -\frac{1}{|\mathcal{D}|}\sum_{u=1}^{|\mathcal{D}|}\log\hat{q}_{u,i_u^{*}},
\label{eq:loss}
\end{equation}
where \(\hat{q}_{u,i_u^{*}}\) is the softmax probability that the network assigns to the optimal beam \(i_u^{*}\).

\subsubsection{Inner-Loop (Support) Update}
Starting from \(\boldsymbol{\theta}^{(0)}_k=\boldsymbol{\theta}\), the classifier head is adapted to \(\mathcal{T}_k\) via \(G_{\mathrm{in}}\) gradient steps on \(\mathcal{S}_k\), while \(\boldsymbol{\Theta}\) and \(\boldsymbol{\Phi}\) remain fixed:
\begin{equation}
\boldsymbol{\theta}^{(g)}_k = \boldsymbol{\theta}^{(g-1)}_k
  - \mu\,\nabla_{\boldsymbol{\theta}^{(g-1)}_k}
    \mathcal{L}_{\mathcal{S}_k}\big(\boldsymbol{\theta}^{(g-1)}_k,\boldsymbol{\Phi};\boldsymbol{\Theta}\big),
\label{eq:inner_update}
\end{equation}
where \(g=1,\dots,G_{\mathrm{in}}\) and \(\mu\) is the inner learning rate.

\subsubsection{Outer-Loop (Query) Update}
The adapted head \(\boldsymbol{\theta}^{(G_{\mathrm{in}})}_k\) is evaluated on \(\mathcal{Q}_k\) of the same task, and \(\boldsymbol{\Phi}\) and \(\boldsymbol{\theta}\) are updated as
\begin{align}
\boldsymbol{\Phi} &\leftarrow \boldsymbol{\Phi} - \alpha\,\nabla_{\boldsymbol{\Phi}}
  \mathcal{L}_{\mathcal{Q}_k}\big(\boldsymbol{\theta}^{(G_{\mathrm{in}})}_k,\boldsymbol{\Phi};\boldsymbol{\Theta}\big),
\label{eq:outer_ss}\\
\boldsymbol{\theta} &\leftarrow \boldsymbol{\theta} - \alpha\,\nabla_{\boldsymbol{\theta}^{(G_{\mathrm{in}})}_k}
  \mathcal{L}_{\mathcal{Q}_k}\big(\boldsymbol{\theta}^{(G_{\mathrm{in}})}_k,\boldsymbol{\Phi};\boldsymbol{\Theta}\big),
\label{eq:outer_head}
\end{align}
where \(\alpha\) is the outer learning rate. Following~\cite{sun2019mtl}, the outer updates are first-order, without
differentiating through the inner loop, and
\(\boldsymbol{\theta}^{(G_{\mathrm{in}})}_k\) is discarded after each episode. In contrast, MAML~\cite{finn2017maml} differentiates through the inner loop, which requires second-order gradients over all network parameters. Unlike~\cite{sun2019mtl}, which discards the pre-trained classifier, MTL-BA retains \(\boldsymbol{\theta}\) since the beam codebook, and hence the label space, is shared across sites.

\subsubsection{Target Environment Adaptation}
After meta-training, \(\boldsymbol{\Theta}\) is kept frozen and \(\boldsymbol{\Phi}\) and \(\boldsymbol{\theta}\) are fine-tuned on a small labeled set \(\mathcal{D}_{\mathrm{ad}}=\{(\mathbf{x}_u,i_u^{*})\}_{u=1}^{N_{\mathrm{ad}}}\) collected at the target site:
\begin{equation}
\{\boldsymbol{\Phi},\boldsymbol{\theta}\} \leftarrow \{\boldsymbol{\Phi},\boldsymbol{\theta}\}
- \eta\,\nabla_{\{\boldsymbol{\Phi},\boldsymbol{\theta}\}}
  \mathcal{L}_{\mathcal{D}_{\mathrm{ad}}}\big(\boldsymbol{\theta},\boldsymbol{\Phi};\boldsymbol{\Theta}\big),
\label{eq:adaptation}
\end{equation}
repeated for \(G_{\mathrm{ad}}\) steps with learning rate \(\eta\). Unlike~\cite{sun2019mtl}, where the SS parameters are frozen at test time, \(\boldsymbol{\Phi}\) is also adapted to the target site. Since \(\boldsymbol{\Theta}\) is frozen, only a small fraction of the network parameters is updated. The meta-training procedure is summarized in Algorithm~\ref{alg:mtl}.
\begin{algorithm}[t]
\caption{MTL-BA Meta-Training}
\label{alg:mtl}
\begin{algorithmic}
\STATE \textbf{Input:} \(\boldsymbol{\Theta}\), \(\boldsymbol{\theta}\), \(\mathcal{B}\), \(\mu\), \(\alpha\), \(G_{\mathrm{in}}\), \(E\), \(T_{\mathrm{ep}}\)
\STATE \textbf{Output:} \(\boldsymbol{\Phi}^{*}\), \(\boldsymbol{\theta}^{*}\)
\STATE Initialize \(\boldsymbol{\phi}^{\gamma}\!\leftarrow\!\mathbf{1}\), \(\boldsymbol{\phi}^{\beta}\!\leftarrow\!\mathbf{0}\)
\FOR{epoch \(e = 1, \dots, E\)}
  \FOR{episode \(t = 1, \dots, T_{\mathrm{ep}}\)}
    \STATE Sample \(\mathcal{T}_k\) from \(b_k\in\mathcal{B}\), draw \(\mathcal{S}_k,\mathcal{Q}_k\)
    \STATE \(\boldsymbol{\theta}^{(0)}_k \leftarrow \boldsymbol{\theta}\)
    \FOR{\(g = 1, \dots, G_{\mathrm{in}}\)}
      \STATE Update \(\boldsymbol{\theta}^{(g)}_k\) via \eqref{eq:inner_update}
    \ENDFOR
    \STATE Update \(\boldsymbol{\Phi}\) and \(\boldsymbol{\theta}\) via \eqref{eq:outer_ss}, \eqref{eq:outer_head}
    \STATE Discard \(\boldsymbol{\theta}^{(G_{\mathrm{in}})}_k\)
  \ENDFOR
\ENDFOR
\RETURN \(\boldsymbol{\Phi}^{*}\leftarrow\boldsymbol{\Phi}\), \(\boldsymbol{\theta}^{*}\leftarrow\boldsymbol{\theta}\)
\end{algorithmic}
\end{algorithm}
\section{Simulation Results}

\begin{figure}[!t]
\centering
\includegraphics[width=0.9\columnwidth]{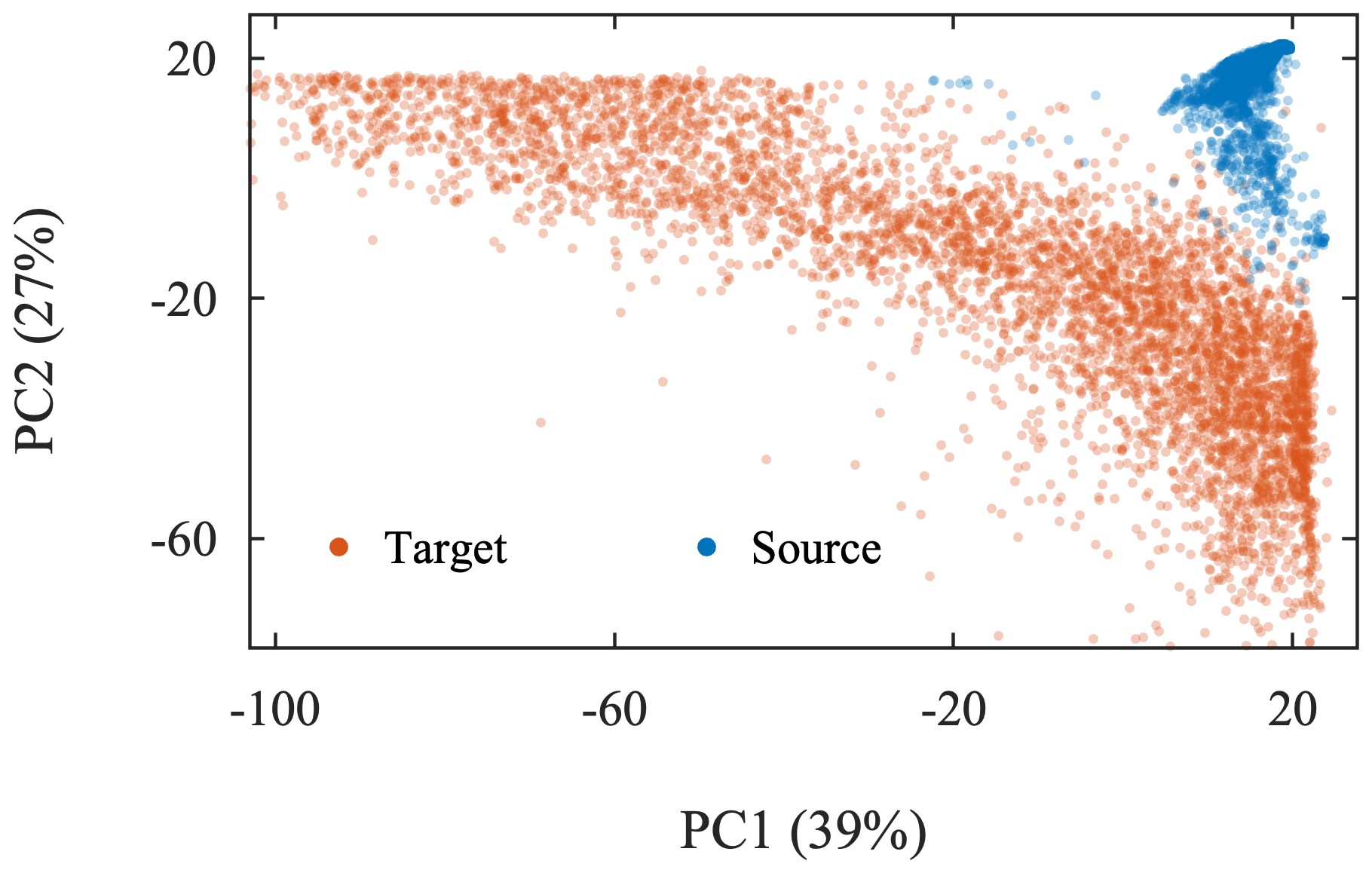}\hspace*{0.08\columnwidth}
\caption{2-D PCA projection of the frozen-backbone features of \(5{,}000\) source (O1\_28) and \(5{,}000\) target (I3\_60) samples.}
\label{fig:shift}
\end{figure}

\begin{figure*}[!t]
\centering
\includegraphics[width=\textwidth]{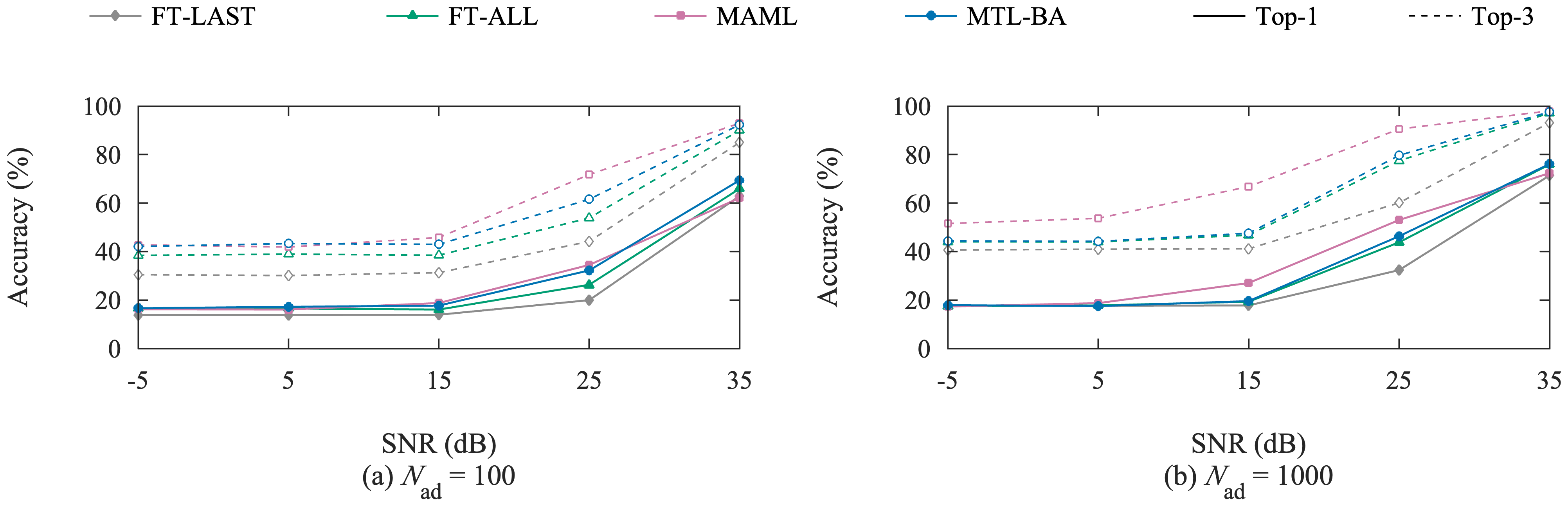}
\caption{Top-1 and Top-3 accuracy versus SNR at the target site, averaged over 5 random seeds.}
\label{fig:acc}
\end{figure*}

\begin{figure*}[!t]
\centering
\includegraphics[width=\textwidth]{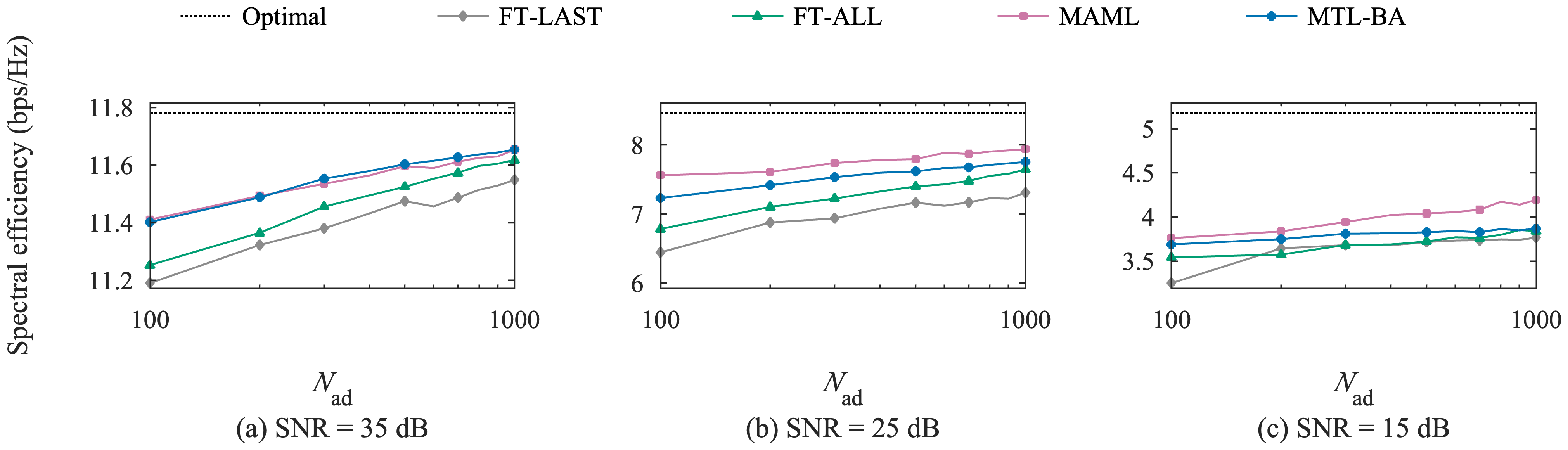}
\caption{Spectral efficiency versus the number of adaptation samples at the target site, averaged over 5 random seeds.}
\label{fig:se}
\end{figure*}

\subsection{Simulation Setup}
\label{subsec:sim_setup}
We use the DeepMIMO ray-tracing dataset~\cite{deepmimo}. BSs~3--13 of the outdoor O1\_28 scenario form the source sites \(\mathcal{B}\), where BSs~4--7 and 9--12 are used for pre-training and meta-training and BSs~3, 8, and 13 for validation. The target site is BS~2 of the indoor I3\_60 scenario. The channel parameters are listed in Table~\ref{tab:channel_params}. For each source BS, the \(50{,}000\) users with the strongest probing measurements are retained, and the target pool contains \(40{,}000\) users. Source measurements include thermal noise with a per-subcarrier power of \(-119.1\)~dBm (noise floor \(-174\)~dBm/Hz, noise figure \(5\)~dB, processing gain \(10\)~dB). At the target site, \(\sigma_n^2\) is varied so that the average optimal-beam SNR ranges from \(-5\) to \(35\)~dB. The narrow-beam codebook \(\mathbf{V}\) is an O-DFT codebook with \(N_V=128\) beams, and the probing codebook is a DFT codebook with \(M_p=32\) beams. The samples \((\mathbf{x}_u,i_u^{*})\) are generated by~\eqref{eq:probe_vec} and~\eqref{eq:opt_beam}.

To quantify the domain shift, all source sites are pooled into a single domain, and \(5{,}000\) samples are drawn from each domain. The samples are passed through the frozen backbone, and the resulting features are projected onto their first two principal components (PC1 and PC2), i.e., the two directions with the largest variance, which together capture \(66\%\) of the total variance. As shown in Fig.~\ref{fig:shift}, the source and target features occupy clearly separated regions. Moreover, a linear classifier trained on these features achieves \(99.8\%\) accuracy on held-out samples when distinguishing the two domains.

\begin{table}[t]
\renewcommand{\arraystretch}{1.0}
\setlength{\tabcolsep}{4pt}
\caption{Channel generation parameters.}
\label{tab:channel_params}
\centering
\begin{tabular}{lcc}
\toprule
\textbf{Parameter} & \textbf{Source} & \textbf{Target} \\ \midrule
Scenario              & O1\_28 (outdoor) & I3\_60 (indoor) \\
Active BSs            & 3--13 & 2 \\
Carrier frequency     & 28~GHz & 60~GHz \\
Transmit power        & \multicolumn{2}{c}{30~dBm} \\
Bandwidth             & \multicolumn{2}{c}{0.5~GHz} \\
BS antennas \((x,y,z)\), spacing & \multicolumn{2}{c}{\((1,32,1)\), \(\lambda/2\)} \\
Subcarriers / sampling / limit   & \multicolumn{2}{c}{512 / 1 / 1} \\
Number of paths \(L\) & \multicolumn{2}{c}{5} \\ \bottomrule
\end{tabular}
\end{table}

\subsection{Model Architecture and Training Configuration}
\label{subsec:dl_arch}

All methods use the CNN in Table~\ref{tab:nn_arch}. The two convolutional
layers and the first fully connected layer (FC1) form the backbone
\(\boldsymbol{\Theta}\), and the second fully connected layer (FC2) is the
classifier head \(\boldsymbol{\theta}\). The CNN is pre-trained on the source
training BSs for \(40\) epochs. MTL-BA updates \(33{,}664\) parameters
(\(768\) SS and \(32{,}896\) head parameters), about \(17\times\) fewer than
the \(570{,}048\) parameters updated by full fine-tuning. For MTL-BA, the
outer updates~\eqref{eq:outer_ss} and~\eqref{eq:outer_head} are implemented
with AdamW. Both MTL-BA and MAML are evaluated on the same validation BSs
during meta-training and trained until the validation performance converges,
requiring \(200\) and \(500\) meta-epochs, respectively. All methods are
adapted with AdamW using mini-batches of \(64\) samples.

\begin{table}[t]
\renewcommand{\arraystretch}{1.0}
\setlength{\tabcolsep}{4pt}
\caption{Neural network architecture and training hyper-parameters.}
\label{tab:nn_arch}
\centering
\begin{tabular}{lcc}
\toprule
\textbf{Parameter} & \textbf{MTL-BA} & \textbf{MAML} \\ \midrule
Conv 1 & \multicolumn{2}{c}{1 $\to$ 64 ch., kernel 3, pad 1} \\
Conv 2 & \multicolumn{2}{c}{64 $\to$ 64 ch., kernel 3, pad 1} \\
FC 1   & \multicolumn{2}{c}{2048 $\to$ 256} \\
FC 2 (head) & \multicolumn{2}{c}{256 $\to$ $N_V$} \\
Activation & \multicolumn{2}{c}{ReLU} \\
Label smoothing (pre-train., adapt.) & \multicolumn{2}{c}{0.05} \\
Episodes per epoch $T_{\mathrm{ep}}$ & \multicolumn{2}{c}{30} \\
Support / query $N_s/N_q$ & \multicolumn{2}{c}{20 / 50} \\
Inner steps $G_{\mathrm{in}}$ & \multicolumn{2}{c}{5} \\
Meta-optimizer & AdamW & Adam \\
Inner LR $\mu$       & $0.1$ & $10^{-2}$ \\
Outer LR $\alpha$    & $10^{-3}$ & $10^{-3}$ \\
Adaptation LR $\eta$ & $10^{-3}$ & $10^{-4}$ \\
Meta-epochs $E$      & 200 & 500 \\ \bottomrule
\end{tabular}
\end{table}

\subsection{Benchmark Methods and Evaluation Metrics}
\label{subsec:benchmarks}
MTL-BA is compared with the following methods, all using the same CNN:
\begin{itemize}
\item \textbf{FT-LAST}~\cite{yuan2021tl_ml}: the pre-trained CNN with only \(\boldsymbol{\theta}\) fine-tuned.
\item \textbf{FT-ALL}: the pre-trained CNN with all parameters fine-tuned.
\item \textbf{AdaBN-style}~\cite{li2018adabn}: since the CNN has no
normalization layers, the per-channel mean and variance of the features
at the SS locations are aligned to those of \(\mathcal{D}_{\mathrm{ad}}\),
followed by fine-tuning \(\boldsymbol{\theta}\).
\item \textbf{LoRA}~\cite{hu2022lora}: a rank-1 update of FC1 is trained together with \(\boldsymbol{\theta}\), resulting in \(35{,}200\) parameters.
\item \textbf{MAML}~\cite{finn2017maml}: all parameters are meta-learned from random initialization with second-order gradients and then fine-tuned.
\end{itemize}
To isolate the contributions of MTL-BA, we also consider SS-FT,
which adapts $\boldsymbol{\Phi}$ and $\boldsymbol{\theta}$ of the
pre-trained CNN without meta-training, and Head-only, which adapts
only $\boldsymbol{\theta}$ of the meta-trained model.

Each method is adapted on \(N_{\mathrm{ad}}\) target samples and evaluated on the remaining target samples. Performance is measured by the Top-\(k\) accuracy, i.e., the fraction of test users whose optimal beam \(i_u^{*}\) is among the \(k\) beams with the highest predicted probabilities. We also evaluate the spectral efficiency achieved by the predicted beam,
defined for user \(u\) as
\begin{equation}
R_u=\log_2\!\left(1+\mathrm{SNR}_{u,\hat{i}_u}\right),
\end{equation}
where \(\hat{i}_u\) is the predicted beam index. The optimal spectral
efficiency is obtained by replacing \(\hat{i}_u\) with \(i_u^\star\).
Mean spectral efficiency is reported in bps/Hz. The training cost is measured by the wall-clock time, GPU-hours, and floating-point operations (FLOPs). Results are averaged over \(5\) random seeds.

\subsection{Performance Evaluation}
\label{subsec:perf}

\subsubsection{Accuracy}

Fig.~\ref{fig:acc} shows the beam prediction accuracy versus SNR. At \(35\)~dB, MTL-BA achieves the highest Top-1 accuracy, \(69.4\%\) and \(76.1\%\) for \(N_{\mathrm{ad}}=100\) and \(1000\), exceeding MAML by \(7.3\) and \(3.7\) points. At \(25\) and \(15\)~dB, MAML achieves the highest accuracy, while MTL-BA remains above FT-ALL and FT-LAST. At \(5\)~dB and below, all methods saturate at \(17\)--\(19\%\) since the probing measurements are dominated by noise.

\subsubsection{Comparison with Other Parameter-Efficient Baselines}
\begin{table}[t]
\renewcommand{\arraystretch}{1.0}
\setlength{\tabcolsep}{4pt}
\caption{Top-1 accuracy (\%) and number of updated parameters.}
\label{tab:comparison}
\centering
\begin{tabular}{lrcccc}
\toprule
& & \multicolumn{2}{c}{\textbf{35 dB}} & \multicolumn{2}{c}{\textbf{25 dB}} \\
\cmidrule(lr){3-4}\cmidrule(lr){5-6}
\textbf{Method} & \textbf{Params} & \(N_{\mathrm{ad}}\!=\!100\) & \(1000\) & \(100\) & \(1000\) \\ \midrule
AdaBN-style    & 32,896  & 52.8 & 71.3 & 21.0 & 43.3 \\
LoRA     & 35,200  & 67.9 & 74.7 & 30.3 & 42.9 \\
MTL-BA   & 33,664  & \textbf{69.4} & \textbf{76.1} & 32.3 & 46.4 \\ \bottomrule
\end{tabular}
\end{table}

Table~\ref{tab:comparison} compares MTL-BA with other parameter-efficient methods. With a comparable number of updated parameters, MTL-BA outperforms AdaBN-style and LoRA in all four settings. Aligning the feature statistics without fine-tuning \(\boldsymbol{\theta}\) yields below \(7\%\) accuracy, indicating that statistics alignment alone cannot compensate for the domain shift.

\subsubsection{Spectral Efficiency}
Fig.~\ref{fig:se} shows the spectral efficiency versus the number of
adaptation samples. At $35$~dB, MTL-BA achieves 96.8\% of the optimal
spectral efficiency with only $N_{\mathrm{ad}}=100$ samples, while
remaining within $0.01$~bps/Hz of MAML. With
$N_{\mathrm{ad}}=1000$, it reaches 98.9\% of the optimal spectral
efficiency and essentially matches MAML. At $25$ and $15$~dB, MAML
achieves higher spectral efficiency, consistent with the accuracy
trends in Fig.~\ref{fig:acc}.

\begin{figure}[!t]
\centering
\includegraphics[width=0.9\columnwidth]{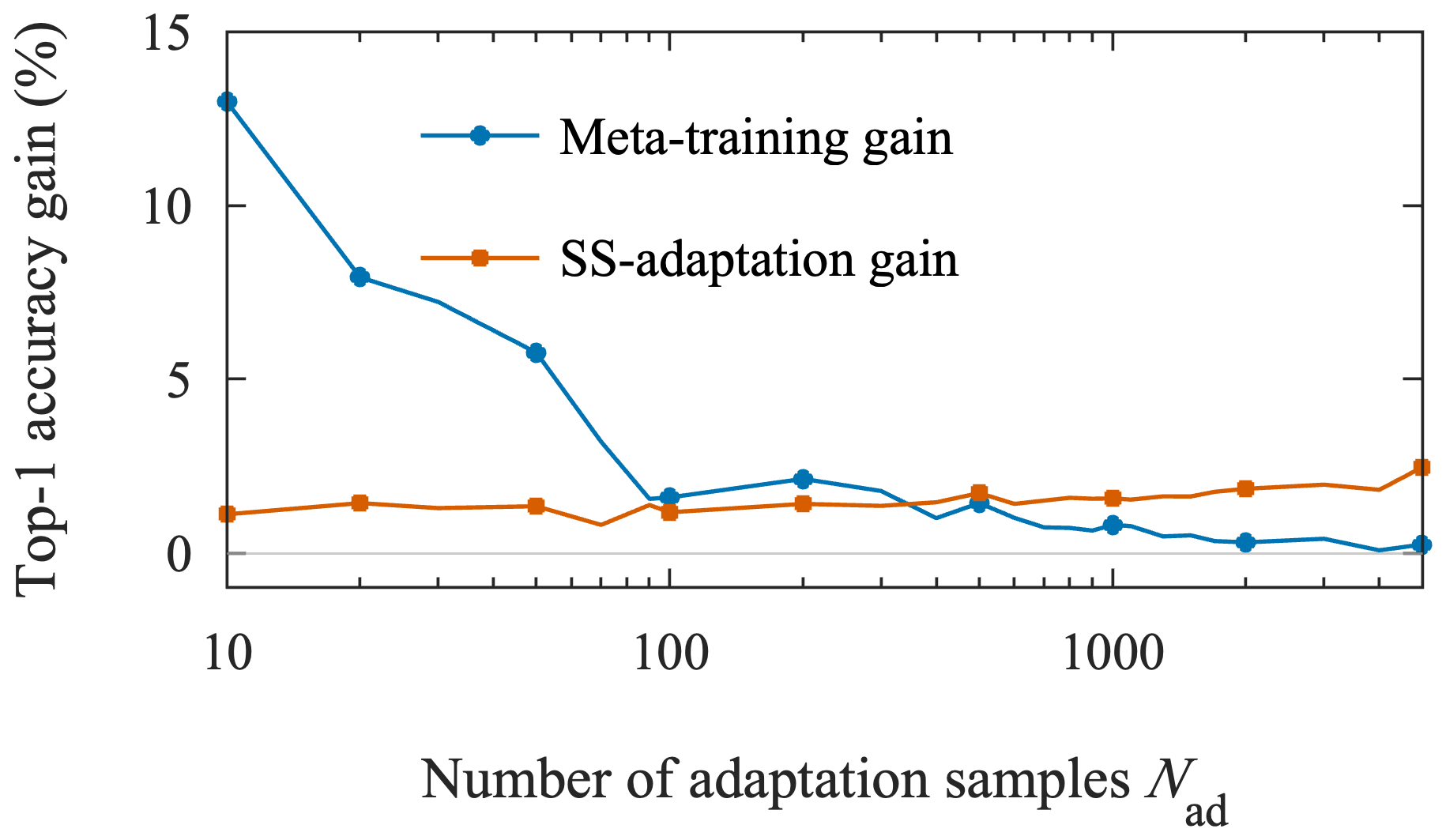}
\caption{Top-1 accuracy gain of MTL-BA over SS-FT (meta-training gain) and over Head-only (SS-adaptation gain) versus \(N_{\mathrm{ad}}\) at \(35\)~dB.}
\label{fig:ablation}
\end{figure}

\subsubsection{Ablation Study}
Fig.~\ref{fig:ablation} separates the contributions of meta-training
and SS adaptation. With \(N_{\mathrm{ad}}=10\), meta-training improves
the accuracy by \(13.0\) points, whereas SS adaptation adds \(1.1\)
points. As \(N_{\mathrm{ad}}\) grows, the meta-training gain decreases
and the SS-adaptation gain dominates, reaching \(1.5\) points at
\(N_{\mathrm{ad}}=1000\). Hence, meta-training provides a better
initialization when samples are scarce, while SS adaptation refines
the frozen features when more adaptation samples are available.

\subsubsection{Training Cost}
\begin{table}[t]
\renewcommand{\arraystretch}{1.0}
\setlength{\tabcolsep}{4pt}
\caption{Training cost per stage on an NVIDIA A100 GPU.}
\label{tab:cost}
\centering
\begin{tabular}{lccc}
\toprule
\textbf{Stage} & \textbf{Time (s)} & \textbf{GPU-h} & \textbf{FLOPs} \\ \midrule
Pre-training          & 128.3 & 0.036 & \(9.2\times10^{13}\) \\
MTL-BA meta-training  & 45.0  & 0.013 & \(2.4\times10^{12}\) \\
MAML meta-training    & 275.7 & 0.077 & \(3.0\times10^{13}\) \\ \bottomrule
\end{tabular}
\end{table}

Table~\ref{tab:cost} reports the training cost. Meta-training of MTL-BA is
\(6.1\times\) faster than MAML and requires \(12.8\times\) fewer FLOPs,
owing to its lower per-episode cost and fewer meta-epochs. Including
pre-training, MTL-BA still requires \(1.6\times\) less time than MAML, but
\(3.1\times\) more FLOPs, since pre-training processes the entire source
dataset. The same pre-trained CNN is shared by all transfer-learning
baselines. During adaptation, MTL-BA requires \(3.9\) MFLOPs per sample
compared with \(5.7\) MFLOPs for FT-ALL and MAML.

\section{Conclusion}

In this letter, we proposed MTL-BA, a meta-transfer learning framework
for adaptive mmWave beam alignment that combines a frozen pre-trained
CNN backbone with lightweight Scale-and-Shift adapters and a classifier
head. Under the considered outdoor-to-indoor domain shift, MTL-BA
achieves high beam prediction accuracy and near-optimal spectral
efficiency at high SNR while adapting substantially fewer parameters
than full fine-tuning and MAML. It also reduces the meta-training cost
compared with MAML. The ablation results show that meta-training is
particularly beneficial when adaptation samples are scarce, whereas SS
adaptation becomes more important as additional target samples become
available. Future work will investigate the integration of MTL-BA into
O-RAN architectures and evaluate its performance on 3GPP-compliant
testbeds and real-world deployments.


\end{document}